\documentclass[
 reprint,
 amsmath,amssymb,
 aps,prl
]{revtex4-1}

\usepackage{graphicx}
\usepackage{dcolumn}
\usepackage{bm}
\usepackage[version=4]{mhchem}

\begin{document}

\preprint{APS/123-QED}

\title{Striped Magnetic Ground State of the Kagom\'{e} Lattice in \ce{Fe4Si2Sn7O16}}

\author{C. D. Ling}
 \email{chris.ling@sydney.edu.au}
 \author{M. C. Allison}
\author{S. Schmid}
\affiliation{%
School of Chemistry, The University of Sydney, Sydney 2006, Australia
}%

\author{M. Avdeev}
\affiliation{
	Australian Centre for Neutron Scattering, Australian Nuclear Science and Technology Organisation, Menai 2234, Australia
}%

\author{J. S. Gardner}
\author{C.-W. Wang}
\affiliation{
	Australian Centre for Neutron Scattering, Australian Nuclear Science and Technology Organisation, Menai 2234, Australia\\
	Neutron Group, National Synchrotron Radiation Research Center, Hsinchu 30077, Taiwan
}%

\author{D. H. Ryan}
\affiliation{%
Physics Department and Centre for the Physics of Materials, McGill University, 3600 University Street, Montreal, Quebec, H3A 2T8, Canada
}%

\author{M. Zbiri}
\affiliation{%
	Institut Laue Langevin, 71 avenue des Martyrs, Grenoble 38042, France
}%

\author{T. S\"{o}hnel}
\affiliation{
	School of Chemical Sciences, University of Auckland, Auckland 1142, New Zealand
}%

\begin{abstract}
	
We have experimentally identified a new magnetic ground state for the kagom\'{e} lattice, in the perfectly hexagonal Fe$^{2+}$ (3d$^6$, $S=2$) compound \ce{Fe4Si2Sn7O16}. Representational symmetry analysis of neutron diffraction data shows that below $T_N=3.5$~K, the spins on $\frac{2}{3}$ of the magnetic ions order into canted antiferromagnetic chains, separated by the remaining $\frac{1}{3}$ which are geometrically frustrated and show no long-range order down to at least $T=0.1$~K. M\"{o}\ss{}bauer spectroscopy confirms that there is no static order on the latter $\frac{1}{3}$ of the magnetic ions -- i.e., they are in a liquid-like rather than a frozen state -- down to at least 1.65~K. A heavily Mn-doped sample Fe$_{1.45}$Mn$_{2.55}$Si$_2$Sn$_7$O$_{16}$ has the same magnetic structure. Although the propagation vector $q=(0,\frac{1}{2},\frac{1}{2})$ breaks hexagonal symmetry, we see no evidence for magnetostriction in the form of a lattice distortion within the resolution of our data. We discuss the relationship to partially frustrated magnetic order on the pyrochlore lattice of \ce{Gd2Ti2O7}, and to theoretical models that predict symmetry breaking ground states for perfect kagom\'{e} lattices. 

\begin{description}
\item[PACS numbers]
75.10.Jm, 75.25.-j, 75.47.Lx, 75.50.Ee, 76.80.+y
\end{description}
\end{abstract}

\maketitle

Ever since the first published consideration of the ground state of a triangular lattice of Ising spins, \cite{wannier} the pursuit of materials with geometrically frustrated magnetic (GFM) lattices has been an important driver in experimental condensed matter physics. \cite{balents} Perfect GFM lattices are proving grounds for a number of predicted exotic states of matter. The most famous of these is the quantum spin liquid (QSL), in which there is effectively no energy barrier between macroscopically degenerate ground states for $S=\frac{1}{2}$ spins, which can therefore continue to fluctuate down to $T=0$~K. \cite{anderson} 

The simplest GFM case is a triangular lattice, followed by the expanded triangular network known as the kagom\'{e} lattice. Undistorted (perfectly hexagonal) magnetic kagom\'{e} lattices are rare -- the most studied examples are naturally occurring minerals or synthetic versions thereof, notably the jarosites $AB_3$\ce{(SO4)2(OH)6} where $B^{3+}$ can be Fe$^{3+}$ ($S=\frac{5}{2}$), \cite{wills1996} Cr$^{3+}$ ($S=\frac{3}{2}$), \cite{inami} or V$^{3+}$ ($S=1$), \cite{papoutsakis} which generally still undergo N\'{e}el ordering due to Dzyaloshinkii-Moriya anisotropy. A series of quinternary oxalates studies by Lhotel \textit{et al.}, \cite{Lhotel_PRL,Lhotel_EPJB} which contain Fe$^{2+}$ ($S=2$) lattices equivalent to kagom\'{e}s (for 1st neighbor interactions only), all freeze into a $q=0$ N\'{e}el state below $\sim3.2$~K. More recently, the Cu$^{2+}$ ($S=\frac{1}{2}$) undistorted kagom\'{e} compound herbertsmithite \ce{ZnCu3(OH)6Cl2} \cite{shores} has attracted a great deal of attention as arguably the most promising QSL candidate so far discovered (for a recent review see ref. \cite{mendels}). 

\begin{figure}[h]
	\includegraphics[width=0.45\textwidth]{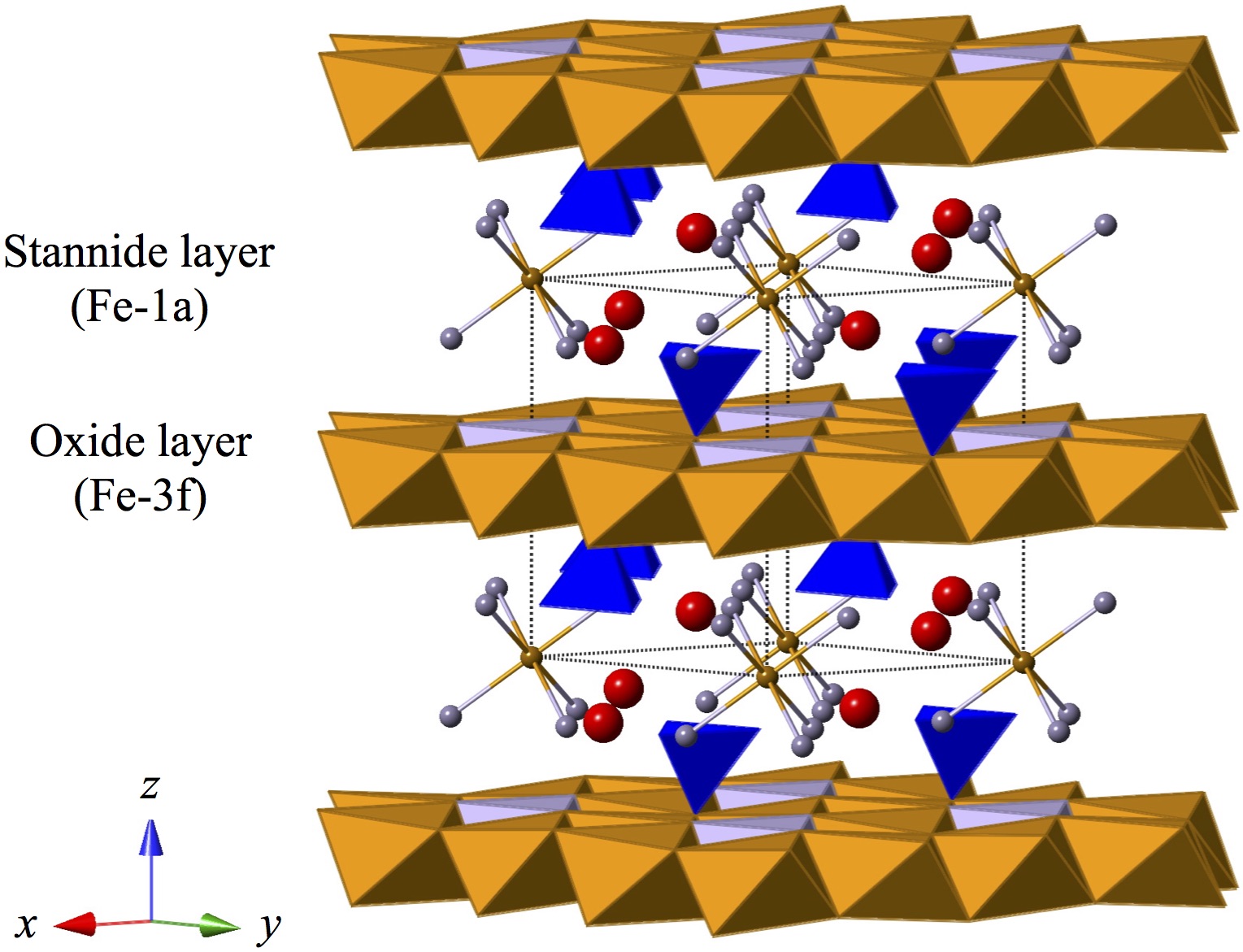}
	\caption{\label{fig:structure}  $P\bar{3}m1$ structure of \ce{Fe4Si2Sn7O16}, showing edge-sharing \ce{FeO6} (gold) and \ce{SnO6} (silver) octahedra in the oxide layer; Fe (gold), Sn (silver) and O (red) atoms in the stannide layer; and \ce{SiO4} tetrahedra (blue) in between.}
\end{figure}

\ce{Fe4Si2Sn7O16} \cite{soehnel} is a synthetic compound that incorporates an undistorted kagom\'{e} lattice of high-spin (HS) Fe$^{2+}$ (3d$^6$, $S=2$) magnetic ions on the 3f Wyckoff sites of its hexagonal (trigonal space group $P\bar{3}m1$, $\#164$) structure. The kagom\'{e} lattice is located in layers of edge-sharing \ce{FeO6} and \ce{SnO6} octahedra (hereafter called the oxide layer), which alternate with layers of oxygen-linked \ce{FeSn6} octahedra (the stannide layer). The layers are separated by \ce{SiO4} tetrahedra (see Fig. \ref{fig:structure}). A triangular lattice of low-spin (LS) Fe$^{2+}$ (3d$^6$, $S = 0$) on the 1a Wyckoff site in the stannide layer is magnetically inactive. 

We recently reported a long-range antiferromagnetic (AFM) N\'{e}el ordering transition in \ce{Fe4Si2Sn7O16} at $T_N=3.0$~K and its Mn-doped (in the oxide layer) analogue Fe$_{1.45}$Mn$_{2.55}$Si$_2$Sn$_7$O$_{16}$ at $T_N=2.5$~K. \cite{allison} There was no evidence of spin-glass behavior or a ferromagnetic (FM) component to the ground state. Given their perfectly hexagonal lattices above $T_N$, the ordered magnetic ground state was expected to be either the conventional $q=0$ or $(\sqrt{3}\times \sqrt{3})$ solution, which preserve hexagonal symmetry. \cite{harris} In the work reported in this Letter, we set out to test this by collecting low-temperature neutron powder diffraction (NPD) and M\"{o}\ss{}bauer spectroscopy data above and below $T_N$. Contrary to expectation, we found that the ground state is a striped AFM structure in which $\frac{2}{3}$ of the magnetic sites are completely ordered, while the other $\frac{1}{3}$ are frustrated and remain completely disordered down to at least 0.1 K. To the best of our knowledge, this state has no precedent experimentally and has not been explicitly predicted theoretically.

New magnetometry data collected for the present study were identical to those in ref. \cite{allison} apart from slightly revised values for \ce{Fe4Si2Sn7O16} of $T_N=3.5$~K, $\mu_{\textrm{eff}}=5.45~\mu_B$ per HS Fe$^{2+}$ and $\theta=-12.7$ K, which corresponds to a modest frustration index \cite{ramirez} $f=|\theta/T_N|=3.6$. \footnote{See Figs. 1--3 of Supplemental Material at [URL will be inserted by publisher]} Note that the orbital angular momentum $L$ is not completely quenched ($\mu_{\textrm{eff spin-only}}=4.90~\mu_B$), as is commonly observed in Fe$^{2+}$ oxides.

High-resolution NPD data were collected on the instrument Echidna \cite{echidna} at the OPAL reactor, Lucas Heights, Australia. Samples were placed in 6 mm diameter vanadium cans using neutrons of wavelength $\lambda = 2.4395$ \AA, over the range $2.75-162^{\circ} 2\theta$ with a step size of $0.125^{\circ} 2\theta$. Low-temperature data were collected to 1.6 K in a cryostat and 0.1 K in a dilution fridge for \ce{Fe4Si2Sn7O16}, and to 1.8 K for Fe$_{1.45}$Mn$_{2.55}$Si$_2$Sn$_7$O$_{16}$. Rietveld-refinements of the nuclear structure above $T_N$ were consistent with our previous work using single-crystal X-ray diffraction \cite{soehnel} and a combination of synchrotron X-ray and neutron powder diffraction.\cite{allison} 

\begin{figure}[h]
	\includegraphics[width=0.45\textwidth]{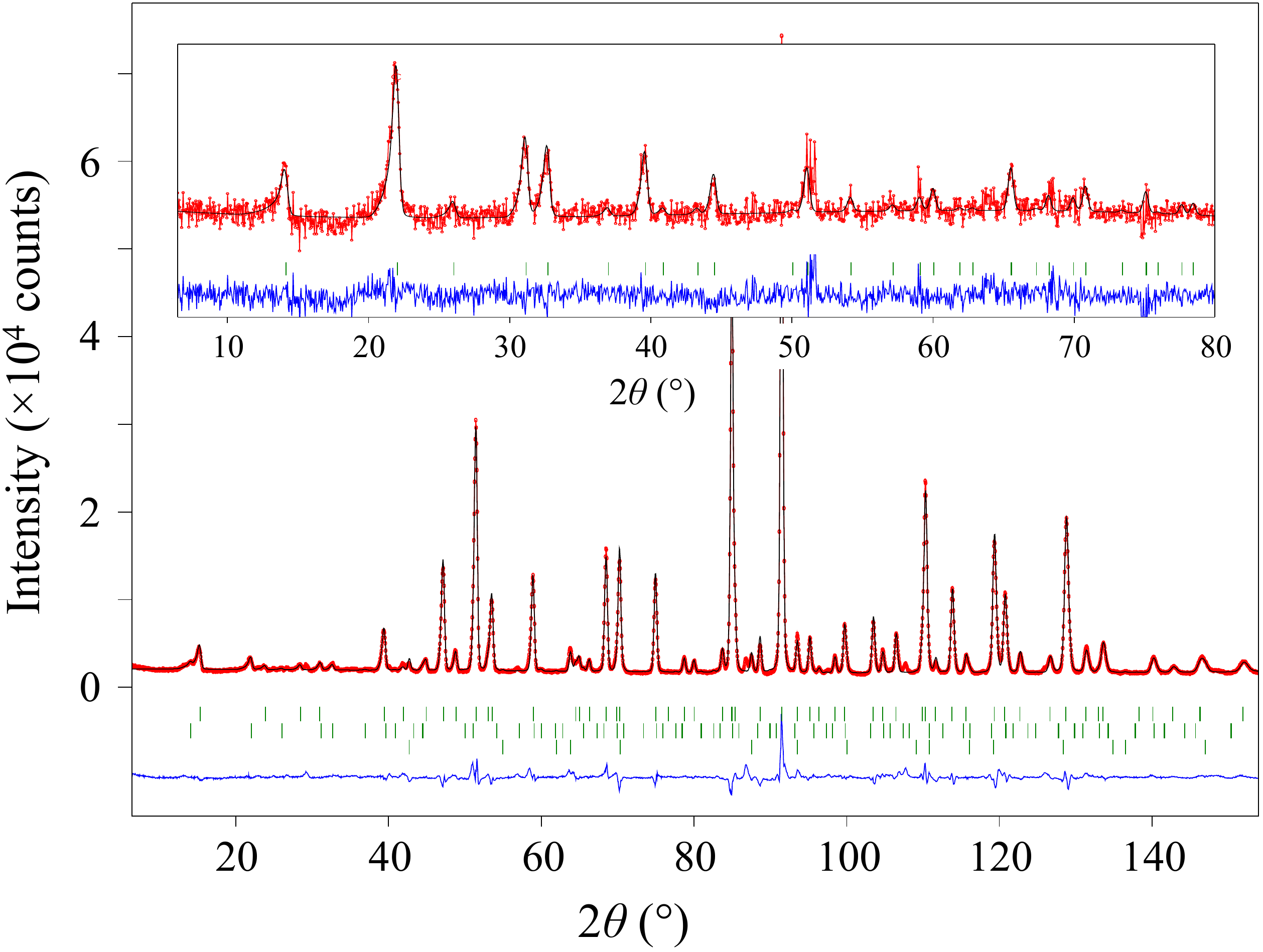}
	\caption{\label{fig:rietveld} (a) Final Rietveld fit (black) to 1.6 K NPD data (red) from \ce{Fe4Si2Sn7O16}, using the $\Gamma_1$ irreducible representation ($R_p=0.121$, $R_{wp}=0.128$). Peak markers from top to bottom correspond to the nuclear and magnetic components of \ce{Fe4Si2Sn7O16}, and a $<1$\%wt \ce{SnO2} impurity. The inset shows the fit to magnetic peaks only (10 K data subtracted from 1.6 K data) at low angles.}
\end{figure}

New low-angle Bragg peaks emerge for \ce{Fe4Si2Sn7O16} at 1.6 K, below $T_N$, indicative of 3D long-range ordered magnetism (Fig. \ref{fig:rietveld}). The same peaks are observed at 0.1 K, i.e., the magnetic structure shows no further change down to at least this temperature; and at 1.8 K for Fe$_{1.45}$Mn$_{2.55}$Si$_2$Sn$_7$O$_{16}$, i.e., the magnetic structure is not fundamentally changed by almost complete substitution of Mn$^{2+}$ for Fe$^{2+}$. \footnote{See Fig. 2 of Supplemental Material at [URL will be inserted by publisher]} All peaks could be indexed to a propagation vector $q=(0,\frac{1}{2},\frac{1}{2})$, which breaks hexagonal symmetry. However, we do not observe any peak splitting or broadening within the resolution of our NPD data, indicating that magnetostriction is very small. 

We solved the magnetic structure by representational symmetry analysis using the BasIreps routine in the program Fullprof. \cite{fullprof} The propagation vector $q=(0,\frac{1}{2},\frac{1}{2})$ acting on the space group $P\bar{3}m1$ splits the magnetic HS Fe$^{2+}$ on the 3f site in the oxide layer into two orbits, Fe-3f(1) and Fe-3f(2), in a 1:2 ratio. The irreducible representations (irreps) for each orbit decompose in terms of two 1D irreps for Fe-3f(1), $\Gamma_{\textrm{mag}}=2\Gamma_2+\Gamma_4$, and another two for Fe-3f(2), $\Gamma_{\textrm{mag}}=3\Gamma_1+3\Gamma_3$. The basis vectors are given in Table 1 of Supplementary Information. 

We first tested these four possible representations independently (i.e., with only the Fe-3f(1) or Fe-3f(2) site magnetically active) by Rietveld refinement against 1.6 K data using Fullprof. We found that the $\Gamma_1$ representation for Fe-3f(2) gave by far the best fit. We tried adding the $\Gamma_2$ and $\Gamma_4$ representations for Fe-3f(1) to the refinement, but in neither case was the fit improved, and the moments on Fe-3f(1) refined to zero within error. The component of the moment on Fe-3f(2) along the $z$ axis also refined to zero within error. The final refinement, for which the fit is shown in Fig. \ref{fig:rietveld}, was therefore carried out using only the $x$ and $y$ axis components of the $\Gamma_1$ representation for Fe-3f(2). This structure is equivalent to the Shubnikov magnetic space group (Opechowski-Guccione setting, $C_{2c}2/m$ \#12.6.71).

\begin{figure}[h]
	\includegraphics[width=0.45\textwidth]{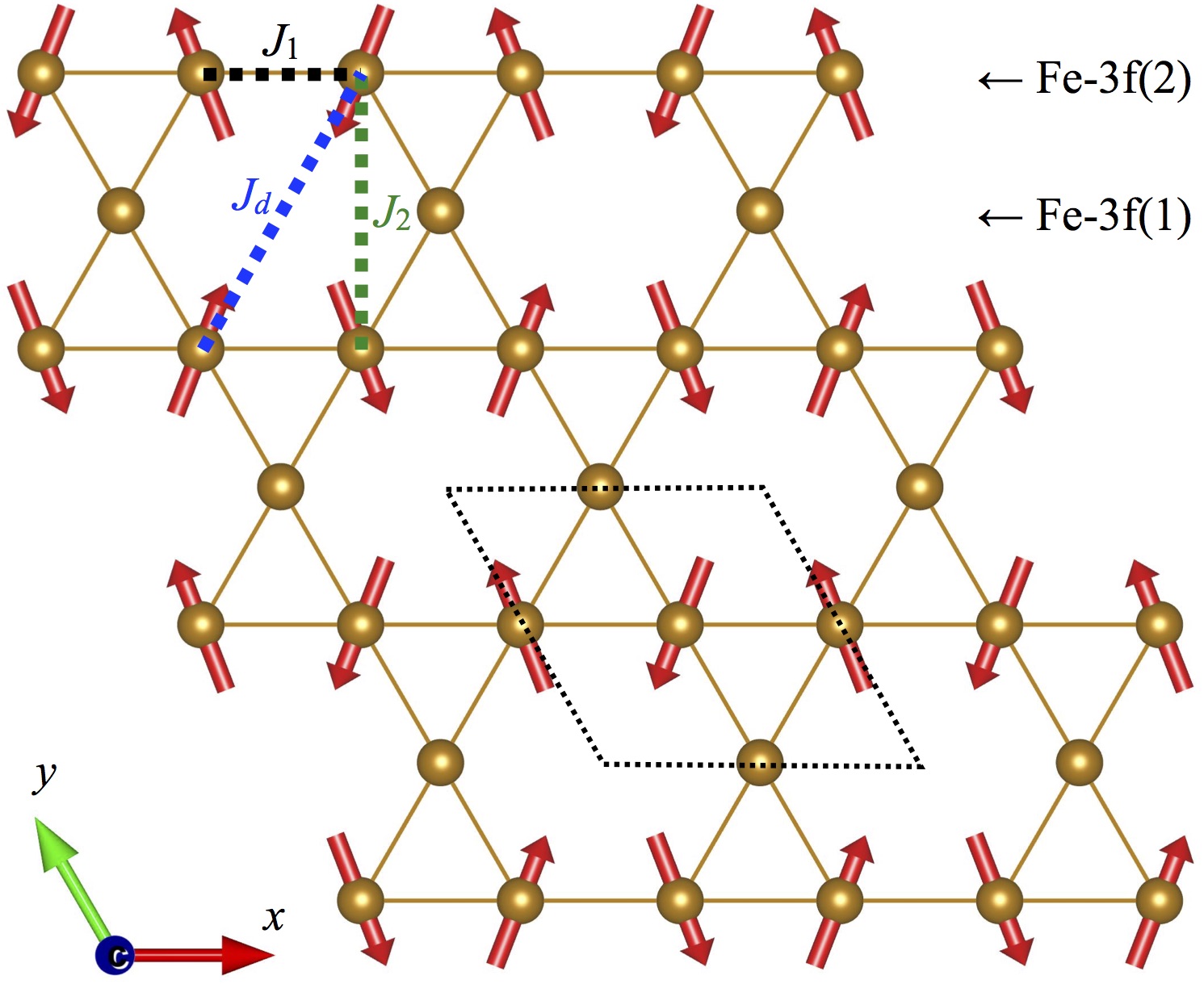}
	\caption{\label{fig:spins} Final refined magnetic structure of \ce{Fe4Si2Sn7O16}, showing the Fe atoms in a single kagom\'{e} layer at $z=\frac{1}{2}$. The nuclear unit cell is shown as dotted black lines. The $z$ axis component refined to zero, and moments on the sites with no vector drawn refined to zero. The moments in the surrounding layers at $z=\pm1c$ are AFM with respect to those shown. Dotted lines show nearest-neighbor ($J_1$, black) second-nearest-neighbor ($J_2$, green) and diagonal ($J_d$, blue) magnetic exchange pathways (see text for details).}
\end{figure}

Fig. \ref{fig:spins} shows the final refined magnetic structure at 1.6 K, with an ordered moment of 2.52(6)~$\mu_B$ ($\mu_x=0.44(6)$, $\mu_y=2.71(5)$~$\mu_B$). The refinement at 0.1 K yielded  increased moment of 3.2(2)~$\mu_B$ ($\mu_x=0.3(2)$, $\mu_y=3.3(2)$~$\mu_B$). The slight reduction compared to the total spin-only moment of 4~$\mu_B$ for HS Fe$^{2+}$ may be an effect of crystal field and spin-orbit coupling. The most striking feature is the absence of long-range magnetic order on the Fe-3f(1) sites, which sit on the kagom\'{e} legs between magnetically ordered rows of Fe-3f(2) sites along the $x$ direction, despite all Fe-3f sites being HS Fe$^{2+}$ (d$^6$, $S=2$) at all temperatures. Note that although the spin orientation in the $x-y$ plane refined robustly, NPD cannot distinguish between the model shown in Fig. \ref{fig:spins} and an alternative version in which the ordered rows of spins are shifted by $\frac{1}{2}a$, to point approximately towards/away from the non-magnetically ordered Fe-3f(1) site rather than towards/away from the center of the hexagon.

It is clear from Fig. \ref{fig:spins} that the Fe-3f(1) site is geometrically frustrated by its four nearest-neighbor Fe-3f(2) sites, regardless of the sign of magnetic exchange with those sites. This is consistent with the zero refined moment for Fe-3f(1), and the fact that because the propagation vector splits the Fe sites into two independent orbits, the molecular field created by one sublattice on the other one is zero, i.e., the 3f(1) magnetic moments can only couple to each other through $J_d$ interactions. However, NPD cannot determine whether Fe-3f(1) is locally ordered but long-range disordered, or completely locally disordered. More importantly, it cannot distinguish between the model shown in Fig. \ref{fig:spins} and a multi-$q$ structure with three arms of the propagation vector star $q_1=(0,\frac{1}{2},\frac{1}{2}$), $q_2=(\frac{1}{2},0,\frac{1}{2}$), $q_3=(\frac{1}{2},\frac{1}{2},\frac{1}{2}$). Such multi-$q$ structures could preserve trigonal symmetry without requiring stripes of disordered Fe$^{2+}$ ions. The anisotropy of Fe$^{2+}$ could be an important parameter in the Hamiltonian to stabilize such a structure. We therefore tested all 1-, 2-, and 3-$q$ symmetry allowed models, with two of the 3-$q$ models (Shubnikov groups $P_C2/m$, $\#$10.8.56 and $P_{2c}\bar{3}m1$, $\#$164.6.1320 in Opechowski-Guccione settings) giving comparable fits to the striped 1-$q$ model discussed above. To resolve the single-$q$ vs. multi-$q$ question, we conducted a M\"{o}\ss{}bauer spectroscopy experiment. 

\begin{figure}[h]
	\includegraphics[width=0.45\textwidth]{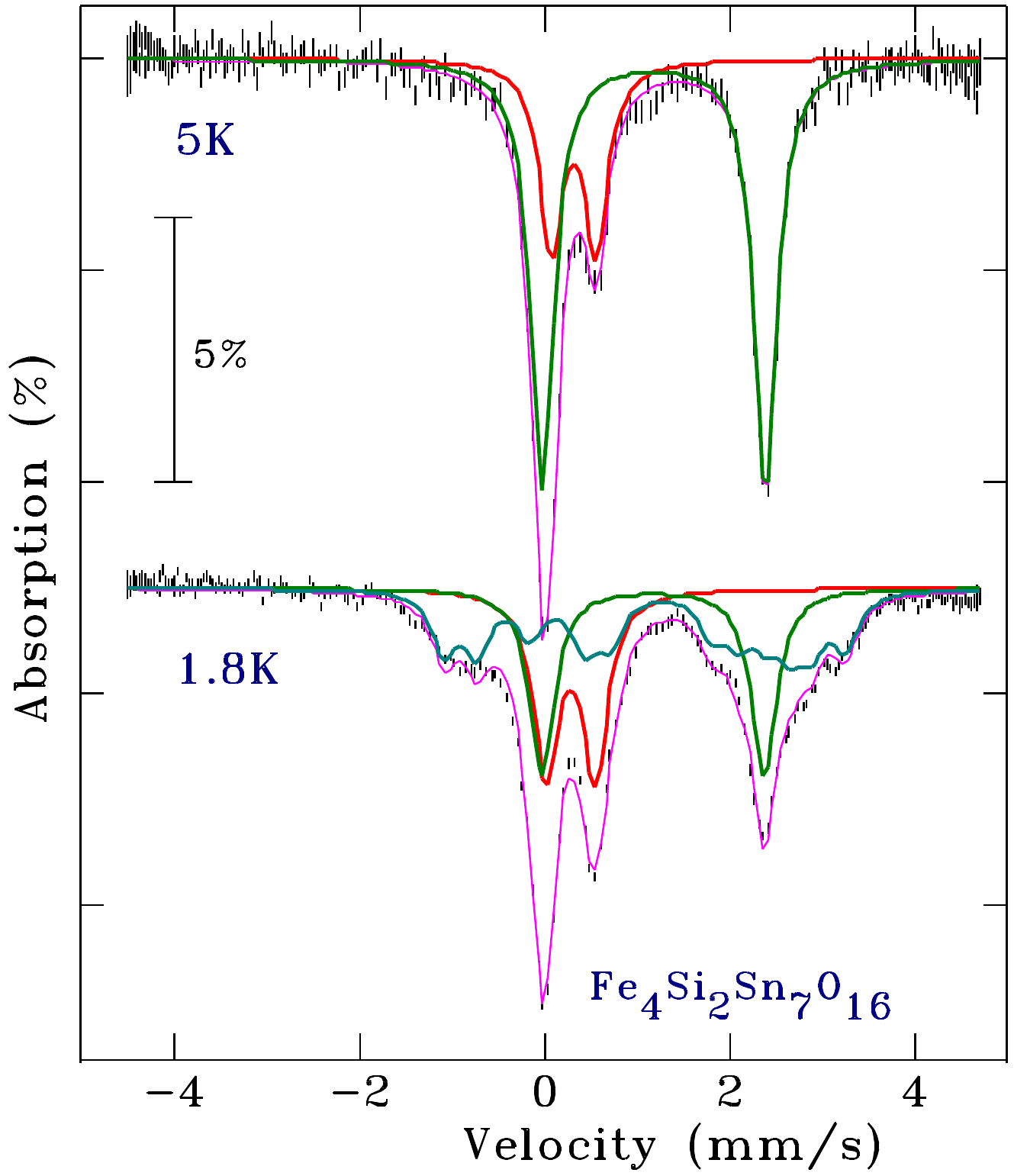}
	\caption{\label{fig:mb} M\"{o}\ss{}bauer spectra above (5~K) and below (1.8~K) $T_N$. For the 5~K pattern, the total fit is shown by the magenta line, as well as the two subspectra from Fe in the 1a (red) and 3f (green) sites. For the 1.8~K pattern the same colours are used except that the contribution from iron in the 3f site is now split into non-magnetic (green) and magnetic (teal).}
\end{figure}

The M\"{o}\ss{}bauer spectra above and below $T_N$ were obtained on a conventional spectrometer operated in sine-mode with both the sample and $^{57}$Co{\bf Rh} source cooled by flowing He gas. The system was calibrated using $\alpha-$Fe metal at room temperature and the spectra were fitted to a sum of Lorentzians with positions and intensities derived from a full solution to the nuclear Hamiltonian. \cite{voyer} The spectrum above $T_N$ at 5~K shown in Fig.~\ref{fig:mb}, which is effectively identical to the data published in ref. \cite{soehnel}, shows that the subspectra from the Fe in the 1a and 3f sites are clearly resolved, being distinct in both spectral area (1:3 as expected from site multiplicities) and hyperfine parameters (as expected from the different spin configurations). The LS Fe$^{2+}$ on the 1a site gives an isomer shift $\delta=0.33(1)$~mm/s with a quadrupole splitting $\Delta=0.48(1)$~mm/s, while the HS Fe$^{2+}$ on the 3f site gives $\delta=1.19(1)$~mm/s and $\Delta = 2.41(1)$~mm/s. Cooling through $T_N$ to 1.8~K leads to remarkably limited changes. The subspectrum from the Fe-1a site is completely unchanged. There is no evidence for magnetic splitting at this site and therefore no magnetic order, consistent with its LS d$^6$ ($S = 0$) configuration. The well-split doublet from the Fe-3f site {\it also} persists unchanged, but its intensity is greatly reduced to equal that of the 1a subspectrum. This non-magnetically ordered subspectrum therefore now reflects only $\frac{1}{3}$ of HS Fe$^{2+}$ on the 3f site, consistent with the Fe-3f(1) site in our single-$q$ striped model. The remaining contribution to the 1.8~K spectrum in Fig.~\ref{fig:mb} comes from the $\frac{2}{3}$ of HS Fe$^{2+}$ 3f sites that do order, consistent with Fe-3f(2) in our striped model. The Fe-3f(2) subspectrum can be fitted assuming the same values for $\delta$ and $\Delta$ as for Fe-3f(1), but with a small additional hyperfine magnetic field ($B_{\textrm{hf}}$). The magnetic component could not be fitted as a single subspectrum, and required further splitting into two equal (within error) area sub-components. This is very weak effect (the fields are only 7.5 T and 4.3 T) that does not change the main result. It may indicate that the moments on the two Fe-3f sub-sites do not make exactly the same angle with the electric field gradient axes in the kagom\'{e} plane, as a result of which they are not truly equivalent, producing some small difference in orbital contribution. 

Crucially, the splitting of the M\"{o}\ss{}bauer spectra is not consistent with any of the multi-$q$ models, including the two mentioned above that fit our NPD fits as well as the striped model does. Finally, the observed hyperfine fields are remarkably small (7.5(3)~T and 4.3(3)~T), and further cooling to 1.65~K did not lead to a significant increase and so these values appear to be close to their $T=0$ limits. The simplest explanation for such small fields is highly dynamic spins, but the dynamic component would have to be very fast ($>100$~MHz) because we not see any line broadening due to slower dynamics.

We note here that while the M\"{o}\ss{}bauer analysis fully and {\it independently} confirms the striped model from NPD in which the Fe-1a site does not order and only $\frac{2}{3}$ of the Fe-3f site orders, it allows us to go further. For moments on a crystallographic site to contribute to a (Bragg) diffraction peak they must be long-range-ordered. However, for a non-zero hyperfine field to be observed in a M\"{o}\ss{}bauer spectrum, the moments need only be static on a time scale of $\sim$0.1~$\mu$s -- they just need to have a non-zero time average over a relatively short time. Thus, our observation that $B_{\textrm{hf}}$ is zero for $\frac{1}{3}$ of the Fe-3f sites means that we can rule out any ``frozen'' random spin configuration that does not give magnetic Bragg peaks in NPD, as well as the multi-$q$ state. There is no static order at $\frac{1}{3}$ of these Fe-3f sites, at least down to 1.65~K. Furthermore, since there are no significant changes in either $\delta$ or $\Delta$, we can also rule out changes in the electronic configuration of some or all of the Fe ions (e.g., a HS $\rightarrow$ LS transition making them non-magnetic).

A further striking aspect of the striped state of \ce{Fe4Si2Sn7O16} is that it breaks hexagonal symmetry. Zorko \textit{et al.} \cite{zorko} recently presented experimental evidence for symmetry breaking in herbertsmithite, but this appears to be related to the presence of significant ($5-8\%$) disorder on its otherwise ``perfect'' kagom\'{e} lattice, which our diffraction data rule out in the present case. A number of theoretical models predict symmetry breaking on $S=\frac{1}{2}$ kagom\'{e} lattices, notably the valence bond crystal (VBC) state \cite{singh} (with the help of magnetoelastic coupling) and the striped spin-liquid crystal state. \cite{clark} However, these models are based on the resonance valence bond (RVB) picture of paired-up $S=\frac{1}{2}$ spins, so their relevance to the present $S=2$ case is unclear. Similarly, a type of striped order was predicted by Ballou \cite{Ballou} in itinerant electron kagom\'{e} systems (the disordered sites being really non-magnetic in this case), but the mechanism should be different to the present insulating case. The magnetic structure of \ce{Fe4Si2Sn7O16} -- and the fact that it is not altered by substituting Mn$^{2+}$ (HS d$^5$, $S=\frac{5}{2}$) for Fe$^{2+}$ (HS d$^6$, $S=2$) -- are important new experimental observations against which to test theoretical models of the large-$S$ kagom\'{e} lattice. In the present case, the fact that the angle between ordered spins along the $x$ axis is very close to $120^{\circ}$ (the value at 0.1 K is $129^{\circ}$) is consistent with the magnetic moments being confined along the two-fold axis, which suggests that magnetocrystalline anisotropy may play an important role.

Preliminary Hubbard-corrected density functional theory (DFT+U) calculations \footnote{We used the projector augmented wave method and GGA-PBE implemented in the Vienna \textit{ab initio} simulations package (VASP), and considered different $U_{\textrm{eff}}=(U-J)$ values to include electron correlation effects stemming from interactions in the Fe[3d] shell \cite{VASP,PAW}. Electronic and magnetic states were stabilized for $U_{\textrm{eff}} > 4$ eV, with a self-consistent field energy convergence breakdown condition set to $10^{-5}$ eV.} for the non-collinear striped ground state of \ce{Fe4Si2Sn7O16} reproduced the zero net moment on the Fe-3f(1) sites. However, the $q=0$ state in the same $1\times 2\times 2$ supercell was still found to be energetically lower. Noting that even the definition of a “ground state” in this system is problematic, dedicated detailed theoretical treatments are clearly required. In this context we note that Iqbal \textit{et al.} \cite{Iqbal} and Gong \textit{et al.} \cite{gong} recently treated the $S=\frac{1}{2}$ kagom\'{e} lattice using high-level renormalization group theory. They identified the dominant magnetic interaction as AFM exchange through the long diagonals of the hexagons, labeled $J_d$ in Fig. \ref{fig:spins}; the ground state then depends on the balance between nearest-neighbor ($J_1$) and second-nearest-neighbor ($J_2$) exchange. For $|J_1|<|J_2|$, they obtain the ``cuboc1'' phase, which is consistent with the vertical components of the spin vectors as shown in Fig. \ref{fig:spins}; while for $|J_1|>|J_2|$, they obtain the ``cuboc2'' phase, which is consistent with the horizontal components. The experimental magnetic structure of \ce{Fe4Si2Sn7O16} can thus be described as a linear combination of cuboc1 and cuboc2. Although this solution was not explicitly predicted for $|J_1|\approx|J_2|$, our experimental case has additional features, notably much bigger spins which make a VBC state highly unlikely, and the presence of $\sim90^{\circ}$ Fe--O--Fe superexchange pathways in addition to Fe--Fe direct exchange. A comparably detailed theoretical study is beyond the scope of the present work, but may represent a productive way forward.  


Finally, we note an intriguing experimental comparison to \ce{Gd2Ti2O7}. The topology of the magnetic Gd$^{3+}$ lattice in this pyrochlore-type compound can be described as four sets of inter-penetrating kagom\'{e} planes, the triangles of which meet to form tetrahedra. Below $T_N=1.1$ K, it adopts the partially ordered ``1-$k$'' structure, in which one of those four sets of kagom\'{e} planes (involving $\frac{3}{4}$ of the spins) is $q=0$ long-range AFM ordered, while the remaining $\frac{1}{4}$ of the spins between those planes remain frustrated. \cite{champion} If the spins in Fig. \ref{fig:spins} are collectively rotated about the $x$-axis, the magnetic structure of \ce{Fe4Si2Sn7O16} becomes equivalent to one of the three other kagom\'{e} planes in \ce{Gd2Ti2O7}, which cut through the frustrated spin. Below $T'=0.7$ K, the frustrated spin in \ce{Gd2Ti2O7} may order weakly into the ``4-$k$'' structure, \cite{stewart} although this has been disputed \cite{Paddison_arXiv} and muon-spin relaxation ($\mu$SR) data show that fluctuations continue down to at least 20 mK. \cite{yaouanc} Future neutron diffuse scattering and/or $\mu$SR experiments at dilution temperatures might therefore provide similar insights into \ce{Fe4Si2Sn7O16}, and the reasons for which it adopts the partially ordered striped state in preference to the fully ordered $q=0$ one. 

The authors received financial support from the Australian Research Council (DP150102863), the School of Chemical Sciences, University of Auckland (FRDF Project 3704173), the Natural Sciences and Engineering Research Council of Canada, and the Fonds Qu\'eb\'ecois de la Recherche sur la Nature et les Technologies.

\bibliography{kagome}

\end{document}